\documentclass[twocolumn,prl,showpacs,preprintnumbers,amsmath,amssymb,superscriptaddress]{revtex4}

\usepackage{graphicx}% Include figure files
\usepackage{dcolumn}% Align table columns on decimal point
\usepackage{bm}% bold math

\begin{document}
\title{What does really mean the ``barrier distribution function''\\
derived from backward angle quasi-elastic scattering}

\author{V.I.~Zagrebaev}
\affiliation{Flerov Laboratory of Nuclear Reactions, JINR, Dubna,
Moscow Region, Russia}

 \maketitle

In a recent Letter \cite{Mitsuoka07} the so called ``barrier
distribution functions'' have been derived from the experimentally
measured quasi-elastic scattering of heavy projectiles ($^{48}$Ti,
$^{54}$Cr, $^{56}$Fe, and $^{64}$Ni) on a $^{208}$Pb target at
backward angles. The authors found that the centroids of the
derived barrier distributions are significantly deviated from the
predicted barrier heights (by 3--10~MeV) toward the low energy
side. The same conclusion was done also in Ref.~\cite{N07} where
the backward angle quasi-elastic scattering was used to derive the
barrier distribution for the reaction $^{86}$Kr+$^{208}$Pb. No
doubts that the measurements are correct. The question arises what
does mean this low-energy shift?

Height of a fusion barrier is an important characteristic of
nucleus-nucleus interaction. Its experimental measurement is
rather difficult, especially for heavy nuclear combinations used
for synthesis of superheavy elements (such as mentioned above). In
this connection, an idea to derive the barrier distribution
function from the quasi-elastic scattering cross section at
backward angles \cite{Timmers95} (which can be measured much
easier) seems very attractive.

In a simple one-dimensional model the idea looks very clear. A
part of incoming flux penetrates the barrier (fusion) with the
transmission probability $T_0(E)$, whereas remaining part reflects
from the barrier (elastic scattering). Height of the barrier, $B$,
can be obtained both from the transmission coefficient (the
derivative $dT_0/dE$ is maximal at $E\approx B$) and from the
reflection coefficient $R_0(E)=1-T_0(E)$, $dR_0/dE=-dT_0/dE$.
Nothing changes if we include in this simple model several
quasi-elastic channels and assume, as before, that fusion is
proportional to the total penetration probability $T^{\rm
fus}(E)=\sum_\nu{T_\nu^{\rm fus}(E)}$ and quasi-elastic scattering
is proportional to the total reflection coefficient $R^{\rm
QE}(E)=\sum_\nu{R_\nu^{\rm QE}(E)}$. Due to unitarity
\begin{equation}
R^{\rm QE}(E)=1-T^{\rm fus}(E) \label{un} \end{equation} and
$dR^{\rm QE}/dE=-dT^{\rm fus}/dE$. This method found its
experimental confirmation for near-barrier collisions of medium
nuclei, such as $^{16}$O+$^{154}$Sm \cite{Timmers95}.

However for heavier combinations, beside quasi-elastic scattering
and fusion, more and more reaction channels appear, mainly deep
inelastic scattering, and instead of Eq.~(\ref{un}) we have
\begin{equation} R^{\rm QE}(E)=1-P^{\rm R}(E), \label{R}
\end{equation} where $P^{\rm R}(E)$ denotes the probability
for {\it all} the reaction channels except the quasi-elasdtic
scattering. The fusion is a largest part of $P^{\rm R}(E)$ for
light projectiles and it is a smaller one for heavy projectiles.
Moreover, in collisions of very heavy nuclei (for example, Xe+Pb)
an interaction potential is repulsive everywhere, there is no
potential pocket and no barrier at all in the strict sense.
Nevertheless, in such collisions the quasi-elastic scattering can
be also measured and relation (\ref{R}) is definitely fulfilled.

This means that the derivative $dR^{\rm QE}/dE=-dP^{\rm R}/dE$
gives us an information not about the ``barrier distribution'' but
about the ``reaction threshold distribution''. For light nuclear
systems these distributions are quite close, but for heavy ones
(when interaction of nuclei becomes more and more complicated and
the potential barrier can disappear at all) they may significantly
differ. The same should hold also for weakly bound neutron rich
projectiles, for which the neutron transfer and break-up reaction
channels play an important role at near-barrier energies.

Obviously, the ``reaction threshold distribution'' should be
shifted to low energy side relative to the ``barrier
distribution'' just because at all incident energies the total
reaction cross section is larger (by definition) than the cross
section of any specific channel (like fusion or capture). For very
heavy nuclear combinations the ``barrier distribution'' loses its
sense at all whereas the ``reaction threshold distribution''
remains its correct physical meaning in accordance with its
definition by Eq.~(\ref{R}). In my opinion it is incorrect to use
different names for the same quantity, $dR^{\rm QE}/dE=-dP^{\rm
R}/dE$, depending on combinations of colliding nuclei.

Note in conclusion that this is not a terminology problem but
important remark, because experiments on measurement of the
backward angle quasi-elastic scattering for deriving the ``barrier
distributions'' for heavy nuclear systems (which may be used for
the production of superheavy elements) are planned to be performed
in several laboratories.

\end{document}